\documentclass[conference,a4paper]{IEEEtran}

\IEEEoverridecommandlockouts

\def\Figs{./}

\usepackage[T1]{fontenc}
\usepackage{ifthen}
\usepackage[cmex10]{amsmath} 

\usepackage[top=0.575in, bottom=0.575in, left=0.55in, right=0.55in]{geometry}

\usepackage[url,hyperrefblack,notheorems,IEEEtran]{research17}
\usepackage{lipsum} 
\usepackage{amsbsy}
\usepackage{booktabs} 
\usepackage{enumitem}
\usepackage{mathtools}
\usepackage{amsmath,amssymb,amsfonts,amsthm} 
\usepackage[lined,boxed,commentsnumbered,linesnumbered,ruled]{algorithm2e}
\usepackage{comment}
\usepackage{tikz} 
\usetikzlibrary{shapes, patterns,decorations.text, decorations.pathreplacing}

\newtheorem{theorem}{\mytheoremname}
\newtheorem{definition}{\mydefinitionname}
\newtheorem{proposition}{\mypropositionname}

\newtheorem{example}{\myexamplename}

\usepackage[capitalize]{cleveref}
\crefname{equation}{\unskip}{\unskip}
\crefname{claim}{Claim}{Claims} 
\usepackage{array}
\usepackage{multirow}
\newcolumntype{C}[1]{>{\centering\arraybackslash}p{#1}}
\allowdisplaybreaks

\newcommand{\rc}[1]{{#1}}

\renewcommand{\vect}[1]{\vectg{#1}} 
\renewcommand{\mat}[1]{\bm{#1}} 
\renewcommand{\vmat}[1]{\bm{#1}} 
\newcommand{\Nat}[1]{\mathbb{N}_{#1}} 
\newcommand{\collect}[1]{\mathscr{#1}} 
\newcommand*{\Scale}[2][4]{\scalebox{#1}{\ensuremath{#2}}} 
\makeatletter
\renewcommand*\env@matrix[1][*\c@MaxMatrixCols c]{%
  \hskip -\arraycolsep
  \let\@ifnextchar\new@ifnextchar
  \array{#1}}
\makeatother

\renewcommand{\HH}{\mathop{}\!\mathsf{H}} 
\renewcommand{\Hb}{\HH_{\tn{b}}} 
\newcommand{\HP}[1]{\HH\left(#1\right)} 
\newcommand{\eHP}[1]{\HH(#1)} 
\newcommand{\bigHP}[1]{\HH\bigl(#1\bigr)}

\newcommand{\HPcond}[2]{\HH\left(#1 \kern0.1em\middle|\kern0.1em #2\right)}
\newcommand{\eHPcond}[2]{\HH(#1 \kern0.1em|\kern0.1em #2)} 
\newcommand{\bigHPcond}[2]{\HH\bigl(#1 \kern-0.1em \bigm| \kern-0.1em#2\bigr)}
\newcommand{\BigHPcond}[2]{\HH\Bigl(#1 \kern-0.1em \Bigm| \kern-0.1em#2\Bigr)}
\renewcommand{\II}{\mathop{}\!\mathsf{I}}  
\newcommand{\MI}[2]{\II\left(#1 \kern0.1em{;}\kern0.1em #2\right)} 
\newcommand{\eMI}[2]{\II(#1 \kern0.1em{;}\kern0.1em #2)} 
\newcommand{\bigMI}[2]{\II\bigl(#1 \kern0.1em{;}\kern0.1em #2\bigr)}
\newcommand{\BigMI}[2]{\II\Bigl(#1 \kern0.1em{;}\kern0.1em #2\Bigr)}
\newcommand{\MIcond}[3]{\II\left(#1 \kern0.1em{;}\kern0.1em #2 \kern0.1em\middle|\kern0.1em #3\right)}
\newcommand{\eMIcond}[3]{\II(#1 \kern0.1em{;}\kern0.1em #2 \kern0.1em|\kern0.1em #3)} 
\newcommand{\bigMIcond}[3]{\II\bigl(#1 \kern0.1em{;}\kern0.1em #2 \kern-0.1em \bigm| \kern-0.1em#3\bigr)}
\newcommand{\BigMIcond}[3]{\II\Bigl(#1 \kern0.1em{;}\kern0.1em #2 \kern-0.1em \Bigm| \kern-0.1em#3\Bigr)}
\renewcommand{\Uniform}[1]{\tn{Uniform}\left(#1\right)}

\renewcommand{\Exp}{\operatorname{\mathbb{E}}}
\newcommand{\lin}[1]{{\color{black} #1}} 

\newcommand{\M}{\mathsf{MI}} 
\newcommand{\bX}{\boldsymbol{X}} 
\newcommand{\cM}{\const{M}} 

\begin{document}
%
\sloppy \title{\textbf{Weakly-Private Information Retrieval}\thanks{This work
    was partially funded by the Research Council of Norway (grant 240985/F20), the Swedish Research Council (grant
    \#2016-04253), and the Israel Science Foundation (grant \#1817/18).}}

 \author{\IEEEauthorblockN{\textbf{Hsuan-Yin Lin}\IEEEauthorrefmark{1}, \textbf{Siddhartha Kumar}\IEEEauthorrefmark{1}, \textbf{Eirik Rosnes}\IEEEauthorrefmark{1}, \textbf{Alexandre Graell i Amat}\IEEEauthorrefmark{2}\IEEEauthorrefmark{1}, and \textbf{Eitan Yaakobi}\IEEEauthorrefmark{3}}
   \IEEEauthorblockA{\IEEEauthorrefmark{1}Simula UiB, N--5008 Bergen, Norway}
  \IEEEauthorblockA{\IEEEauthorrefmark{2}Department of Electrical Engineering, Chalmers University of Technology,
     SE--41296 Gothenburg, Sweden}
       \IEEEauthorblockA{\IEEEauthorrefmark{3}Department of Computer Science, Technion --- Israel Institute of Technology, Haifa, 3200009 Israel}
     }

\thispagestyle{empty}
\pagestyle{empty}

\maketitle

\begin{abstract}
  \emph{Private information retrieval} (\emph{PIR}) protocols make it possible to retrieve a file from a database
  without disclosing any information about the identity of the file being retrieved. These protocols have been
  rigorously explored from an information-theoretic perspective in recent years. While existing protocols strictly
  impose that no information is leaked on the file's identity, this work initiates the study of the tradeoffs that can
  be achieved by relaxing the requirement of perfect privacy. In case the user is willing to leak some information on
  the identity of the retrieved file, we study how the PIR rate, as well as the upload cost and access complexity, can
  be improved. For the particular case of replicated servers, we propose two \emph{weakly-private} information retrieval schemes based on two recent
  PIR protocols and a family of schemes based on partitioning. Lastly, we compare the performance of the proposed
  schemes.
\end{abstract}

\section{Introduction}
\label{sec:introduction}

In 1995 Chor \emph{et al.} introduced the notion of private information retrieval (PIR)
\cite{ChorGoldreichKushilevitzSudan95_1}. A PIR scheme allows a user to privately retrieve an arbitrary file from a
database that is stored in multiple noncolluding servers without revealing any information about the requested file
index to any server. The efficiency of a PIR scheme is usually measured in terms of the communication load, which is the
sum of the number of uploaded and downloaded bits for retrieval of a single file. It has been extensively studied how it
is possible to reduce the communication load using several copies of the
database~\cite{BeimelIshaiKushilevitRaymond02_1,DvirGopi16_1,Yekhanin08_1}. To achieve a more efficient PIR scheme, PIR
protocols have also been considered jointly with \emph{coded distributed storage systems} (DSSs), where the data is
encoded by a linear code to store the files on $n$ servers in a distributed
manner~\cite{ChanHoYamamoto15_1,FazeilVardyYaakobi15_1}.

Recently, there has been a renewed interest to study the PIR problem from an information-theoretic formulation
\cite{ChanHoYamamoto15_1,SunJafar17_1,TajeddineGnilkeElRouayheb18_1}. Under this setting, the file size is assumed to be
arbitrarily large, and hence the upload cost can be ignored compared to the download cost. This then defines the
\emph{PIR rate} for a PIR scheme, which is equal to the amount of information retrieved per downloaded symbol. Recently,
Sun and Jafar derived the optimal achievable PIR rate, the so-called \emph{PIR capacity}, for the classical PIR model of
replicated servers \cite{SunJafar17_1, SunJafar18_2}. Since then, several works have extended the results on PIR 
 to different setups, e.g., coded
DSSs~\cite{BanawanUlukus18_1,KumarRosnesGraellAmat17_1,KumarLinRosnesGraellAmat19_1app}, colluding
servers~\cite{SunJafar18_2,KumarLinRosnesGraellAmat19_1app},
and different figures of merit such as the access complexity~\cite{ZhangYaakobiEtzionSchwartz19_1app}.

The PIR model has also been extended in different interesting directions, for example PIR with side
information~\cite{KadheGarciaHeidarzadehElRouayhebSprintson17_1} and more.
All of the aforementioned models impose the strict requirement of perfect privacy, i.e., no information leakage.
However, this assumption is quite restrictive and may be relaxed for practical applications.
How to quantify the amount of sensitive information leaked from different privacy-enhancing technologies has been
studied massively in the computer science society \cite{WagnerEckhoff18_1}. Therein, many information-theoretic
privacy leakage metrics have been proposed, e.g., mutual information (MI) and worst-case entropy measures
\cite{KopfBasin07_1}.

\lin{This paper takes a first step towards another parameter of the PIR framework, namely the information leakage.
  In particular, the goal of this paper is to study the tradeoffs of the different parameters of PIR protocols, such as
  the rate, upload cost, and access complexity, while the user is willing to leak some information on the identity of
  the retrieved file. We refer to such a scenario as \emph{weakly-private} information retrieval (WPIR). Although
  related, our model is different from the one considered by Toledo \emph{et al.} \cite{ToledoDanezisGoldberg16_1} in the computer
  science literature, where a modified metric based on \emph{differential privacy} that relies on a particular
  scenario between an adversary and a number of users, is considered.} In several scenarios, leaking part of the
information of the retrieved file's identity is legitimate as long as there is still enough ambiguity on the file to
meet the privacy requirement specified by the user. For example, the user may be willing to share with the servers that
the file is a movie (and not a book or other forms of files), or only the movie's genre, however the identity of the
movie should be kept private.

The rest of the paper is organized as follows. Section~\ref{sec:setup-preliminaries} presents the notation, 
definitions, and  system model used throughout the paper. In Section~\ref{sec:partition-schemes}, a basic solution for WPIR is
presented in which the database is partitioned into several partitions and the user is willing to expose only the
partition that the requested file belongs to. In Section~\ref{sec:Scheme1}, we propose a WPIR scheme for replicated
databases building upon a PIR protocol recently introduced in~\cite{TianSunChen18_1sub} and study its tradeoffs between
different parameters while relaxing the privacy constraint. A second WPIR scheme is presented \rc{in}
Section~\ref{sec:Scheme2}, based on the PIR scheme from~\cite{{KumarLinRosnesGraellAmat19_1app}}. Lastly,
Section~\ref{sec:numerical-results} presents numerical results and compares the schemes studied in the paper
with respect to rate, \rc{upload cost, and access complexity}.

\section{Preliminaries}
\label{sec:setup-preliminaries}

\subsection{Notation}
\label{sec:notation-definitions}

We denote by $\Naturals$ the set of all positive integers, $[a]\eqdef\{1,2,\ldots,a\}$, and
$[a:b]\eqdef\{a,a+1,\ldots,b\}$ for $a,b\in\{0\}\cup\Naturals$, $a \leq b$. Vectors are denoted by bold letters, random
variables (RVs) (either scalar or vector) by uppercase letters, and sets by calligraphic uppercase letters, e.g.,
$\vect{x}$, $X$, and $\set{X}$, respectively.
For a given index set $\set{S}$, we write $X^\set{S}$ and $Y_\set{S}$ to represent
$\bigl\{X^{(m)}\colon m\in\set{S}\bigr\}$ and $\bigl\{Y_l\colon l\in\set{S}\bigr\}$, respectively.
$X\indep Y$ means that the two RVs $X$ and $Y$ are independent. $\E[X]{\cdot}$ denotes expectation over \rc{the} RV
$X$. $X\sim\Bernoulli{p}$ denotes a Bernoulli-distributed RV with $\Prv{X=1}=p=1-\Prv{X=0}$ and $X\sim\Uniform{\set{S}}$
a uniformly-distributed RV over the set $\set{S}$.
$\trans{(\cdot)}$ denotes the transpose of its argument. The Hamming weight of a binary vector $\vect{x}$ is denoted by
$w_\mathsf{H}(\vect{x})$ and the inner product of $\vect{x}$ and $\vect{y}$ is denoted by
$\langle\vect{x},\vect{y}\rangle$. $\HP{X}$ represents the entropy $X$ and $\eMI{X}{Y}$ the MI between $X$ and $Y$. The
Galois field with $q$ elements is denoted by $\GF(q)$.

\subsection{System Model}
\label{sec:system-model}

We consider a DSS with $n$ noncolluding replicated servers, each storing $\const{M}$ independent files
$\vmat{X}^{(1)},\ldots,\vmat{X}^{(\const{M})}$, where each file
$\vmat{X}^{(m)}=\trans{\bigl(X_1^{(m)},\ldots,X_\beta^{(m)}\bigr)}$, $m\in [\const{M}]$, can be seen as a
$\beta\times 1$ vector over $\GF(q)$. Assume that each element of $\vmat{X}^{(m)}$ is chosen independently and uniformly
at random from $\GF(q)$. Thus, in $q$-ary units, we have $\bigHP{\vmat{X}^{(m)}}=\beta$, $\forall\,m\in [\const{M}]$.


In information retrieval (IR), a user wishes to efficiently retrieve one of the $\const{M}$ files stored in the
replicated DSS. Similar to the detailed
mathematical description in \cite{TianSunChen18_1sub}, we assume that the requested file index $M$ is a RV
and $M\sim\Uniform{[\const{M}]}$. We give the following definition \rc{of IR schemes.}
\begin{definition}
  \label{Def:Mn-IRscheme}
  An $(\const{M},n)$ IR scheme $\collect{C}$ for a DSS with $n$  servers storing $\const{M}$ files
  consists of
  \begin{itemize}
  \item a global random strategy $\vmat{S}$, whose alphabet is $\set{S}$,
  \item $n$ query-encoding functions $\phi_l$, $l\in [n]$, that generate $n$ queries $\vmat{Q}_l=\phi_l(M,\vmat{S})$ with
    alphabet $\set{Q}_l$, where query $\vmat{Q}_l$ is sent to server $l$\rc{,}
  \item $n$ answer functions $\varphi_l$ that return the answers
    $\vmat{A}_l=\varphi_l(\vmat{Q}_l,\vmat{X}^{[\const{M}]})$, with alphabet  $\set{A}$ for all
    $l\in [n]$,
  \item $n$ answer-length functions $L_l(\vmat{A}_l)$, with range $\{0\}\cup\Nat{}$, that define the length of the answers,
  \item $n$ access-number functions $\delta_l(\vmat{Q}_l)$, with range $\{0\}\cup\Nat{}$, that define the number of
    symbols accessed by $\vmat{Q}_l$.
  \end{itemize}
  This scheme should satisfy the condition of \emph{perfect retrievability},
  \begin{IEEEeqnarray}{rCl}
    \bigHPcond{\vmat{X}^{(M)}}{\vmat{A}_{[n]},\vmat{Q}_{[n]},M}=0.
    \label{eq:retrievability}
  \end{IEEEeqnarray}
\end{definition}

Note that a PIR scheme is an $(\const{M},n)$ IR scheme that satisfies full privacy
for all servers, i.e., for every $m,m'\in[\const{M}]$ with $m\neq m'$, the condition
\begin{IEEEeqnarray}{c}
  \label{eq:strong-privacy}
  \Prvcond{\vmat{Q}_l=\mat{q}_l}{M=m}=\Prvcond{\vmat{Q}_l=\mat{q}_l}{M=m'}
\end{IEEEeqnarray}
holds for all $\mat{q}_l\in\set{Q}_l$, $l\in [n]$. The privacy constraint \eqref{eq:strong-privacy} is equivalent to the
statement that
$M\indep \vmat{Q}_l$. We denote by $\vmat{Q}_l^{(m)}$ the query sent to server $l$ if file $\bX^{(m)}$ is requested, which
is a RV with probability mass function (PMF) $P_{\vmat{Q}_l^{(m)}}(\mat{q}_l)\eqdef\Prvcond{\vmat{Q}_l=\mat{q}_l}{M=m}$.

We refer to an $(\const{M},n)$ IR scheme that does not satisfy \eqref{eq:strong-privacy} as a \emph{WPIR scheme}, as
opposed to a PIR scheme that leaks no information.


\subsection{Metrics of Information Leakage}
\label{sec:metrics_information-leakage}


We consider both the MI and the worst-case information leakage (WIL) \cite{KopfBasin07_1} between
$M$ and $\vmat{Q}_l$ to define suitable measures of information leakage for an IR scheme. For the former, we use the
following proposition to motivate the definition of information leakage for an $(\const{M},n)$ IR scheme.
\begin{proposition}[Time-Sharing Principle for \rc{the} MI Metric]
  \label{prop:time-sharing_MI}
  Consider an $(\const{M},n)$ IR scheme $\collect{C}$, where the leakage of the $l$-th server is defined as
  $\eMI{M}{\vmat{Q}_l}$, $l\in[n]$. Then, there exists an $(\const{M},n)$ IR scheme $\bar{\collect{C}}$ with leakage
  $\bar{\rho}\eqdef\frac{1}{n}\sum_{l\in[n]}\eMI{M}{\vmat{Q}_l}$ for every server.
\end{proposition}
\lin{Proposition~\ref{prop:time-sharing_MI} indicates that we can  obtain an $(\const{M},n)$ IR scheme with equal MI leakage at each server by
  cyclically shifting the servers' queries of an existing $(\const{M},n)$ IR scheme $n$ times.}

Hence, to characterize the overall leakage of a given $(\const{M}, n)$ IR scheme \rc{$\collect{C}$} in terms of MI, we
consider the information leakage metric
$\rho^{(\mathsf{MI})}(\collect{C})\eqdef\frac{1}{n}\sum_{l\in [n]}\eMI{M}{\vmat{Q}_l}$.



\begin{definition}
  \label{def:worst-case-MI-leakage}
  The WIL of the $l$-th server is defined as
  $\II^\mathsf{worst}(M;\vmat{Q}_l)=\eHP{M}-\min_{\mat{q}_l\in\set{Q}_l}\eHPcond{M}{\vmat{Q}_l=\mat{q}_l}$. The overall
  WIL of a given $(\const{M},n)$ IR scheme $\collect{C}$ is then given as
  $\rho^{(\mathsf{WIL})}(\collect{C})\eqdef\max_{l\in [n]}\II^\mathsf{worst}(M;\vmat{Q}_l)$.

\end{definition}

\subsection{Download Cost, Rate, Upload Cost, and Access Complexity of an $(\const{M},n)$ IR Scheme}
\label{sec:achievable-rate_IR}

The download cost of an IR scheme $\collect{C}$, denoted by $\const{D}(\collect{C})$, is defined as the expected number
of downloaded symbols among all servers for the retrieval of a single file,
\begin{IEEEeqnarray*}{rCl}
  \const{D}(\collect{C})\eqdef\sum_{l=1}^n\E[\vmat{Q}_l]{L_l(\vmat{A}_l)}=
  \frac{1}{\const{M}}\sum_{m=1}^{\const{M}}\sum_{l=1}^n\E[\vmat{Q}_l^{(m)}]{L_l(\vmat{A}_l)}.
\end{IEEEeqnarray*}
Accordingly, the IR rate of an IR scheme $\collect{C}$ is defined as
$\const{R}(\collect{C})\eqdef\frac{\beta}{\const{D}(\collect{C})}$. The upload cost $\const{U}(\collect{C})$ of an IR
scheme \rc{$\collect{C}$} is defined as the sum of the entropies of the queries $\vmat{Q}_{[n]}$,
\begin{IEEEeqnarray*}{c}
  \const{U}(\collect{C})\eqdef\sum_{l=1}^n\eHP{\vmat{Q}_l}.\label{eq:def_upload}
\end{IEEEeqnarray*}
Moreover, the access complexity $\Delta(\collect{C})$ of an IR scheme \rc{$\collect{C}$} is defined as the expected
number of accessed symbols among all servers for the retrieval of a single file,
\begin{IEEEeqnarray*}{c}
  \Delta(\collect{C})\eqdef\sum_{l=1}^n\E[\vmat{Q}_l]{\delta_l(\vmat{Q}_l)}
  =\frac{1}{\const{M}}\sum_{m=1}^{\const{M}}\sum_{l=1}^n\E[\vmat{Q}_l^{(m)}]{\delta_l(\vmat{Q}_l)}.
\end{IEEEeqnarray*}




An achievable $4$-tuple \rc{of an IR scheme} is defined as follows.
\begin{definition}
  \label{def:tuple_IR}
  Consider a DSS with $n$ noncolluding servers storing $\const{M}$ files. A $4$-tuple $(\const{R},\const{U},\Delta,\rho)$
  is said to be \emph{achievable} with information leakage metric $\rho^{(\cdot)}$ if there exists an $(\const{M},n)$ IR
  scheme $\collect{C}$ such that $\const{R}(\collect{C})=\const{R}$, $\const{U}(\collect{C})=\const{U}$,
  $\Delta(\collect{C})=\Delta$, and $\rho^{(\cdot)}(\collect{C})=\rho$.
\end{definition}

We remark that a PIR scheme is equivalent to an $(\const{M},n)$ IR scheme with $\rho^{(\cdot)}=0$. It was shown in
\cite{SunJafar17_1} that for $n$ noncolluding replicated servers and for a given number of files $\const{M}$, the PIR
capacity, denoted by $\const{C}_{\const{M},n}$, is
$\const{C}_{\const{M},n}=\inv{\bigl(1+1/n+\cdots+1/n^{\const{M}-1}\bigr)}$.

\section{Partition WPIR Scheme}
\label{sec:partition-schemes}

A simple approach for the construction of WPIR schemes is to first partition the database into $\eta$ equally-sized
partitions, each consisting of $\const{M}/\eta$ files\footnote{While it is not necessary that each partition has an
  equal number of files, for simplicity in this paper we make this assumption.}, and then use a given
$(\const{M}/\eta,n)$ IR scheme to retrieve a file from the corresponding partition. Obviously, the resulting scheme is
not a PIR scheme, since the servers gain the knowledge of which partition the requested file belongs to. In this
section, we pursue this approach to construct an $(\const{M},n)$ IR scheme building on a given $(\const{M}/\eta,n)$ IR
scheme as a subscheme.

The partition $(\const{M},n)$ IR scheme is formally described as follows.  Assume the requested file $\vmat{X}^{(m)}$
belongs to the $j$-th partition, where $j\in[\eta]$.
Then, \rc{the query} $\vmat{Q}_l$ is constructed as
\begin{IEEEeqnarray}{c}
  \vmat{Q}_l=\bigl(\tilde{\vmat{Q}}_l,j\bigr)\in\tilde{\set{Q}}_l\times [\eta],\quad l\in[n],
  \label{eq:queries_partition-scheme}
\end{IEEEeqnarray}
where $\tilde{\vmat{Q}}_l$ is the query of an \rc{existing} $(\const{M}/\eta,n)$ IR scheme $\tilde{\collect{C}}$.

The following theorem states the achievable $4$-tuple of the partition scheme.
\begin{theorem}
  \label{thm:partition-schemes}
  Consider a DSS with $n$ noncolluding servers storing $\const{M}$ files, and let $\tilde{\collect{C}}$ be an
  $(\const{M}/\eta,n)$ IR scheme with achievable 4-tuple
  $\bigl(\tilde{\const{R}},\tilde{\const{U}},\tilde{\Delta}, \tilde{\rho}^{(\cdot)}\bigr)$. Then, the $4$-tuple
  \begin{IEEEeqnarray}{rCl}
    \IEEEeqnarraymulticol{3}{l}{%
      \bigl(\const{R}(\collect{C}),\const{U}(\collect{C}),\Delta(\collect{C}),\rho^{(\cdot)}(\collect{C})\bigr)}
    \nonumber\\*\quad%
    & = &
    \bigl(\tilde{\const{R}},\tilde{\const{U}}+n\log_2{\eta},\tilde{\Delta},\tilde{\rho}^{(\cdot)}+\log_2{\eta}\bigr)
    \label{eq:tuple_partition-scheme}
  \end{IEEEeqnarray}
  is achievable by the $(\const{M},n)$ partition scheme $\collect{C}$ constructed from $\tilde{\collect{C}}$ as
  described in \eqref{eq:queries_partition-scheme}.
\end{theorem}

Since a PIR scheme is also an IR scheme, this simple approach for the construction of WPIR schemes can also be adapted
to use one of the existing $(\const{M}/\eta,n)$ PIR schemes in the literature as a subscheme. We refer to the partition
scheme that uses a PIR scheme as the underlying subscheme and the query generation
in~\eqref{eq:queries_partition-scheme} as a \emph{basic scheme} and denote it by $\collect{C}^{\mathsf{basic}}$ (it
gives the $4$-tuple as in~\eqref{eq:tuple_partition-scheme} with $\tilde{\rho}^{(\cdot)}=0$). In
Section~\ref{sec:partition-SchemeA}, we will present another partition WPIR scheme $\collect{C}'$ based on our proposed
IR scheme.

\section{$(\const{M},n)$ Scheme~1}
\label{sec:Scheme1}

In \cite[Sec.~3.2]{TianSunChen18_1sub}, a PIR scheme that achieves both the minimum upload and download \rc{costs} was
proposed. The queries $\vmat{Q}_{[n]}$ of the scheme in \cite[Sec.~3.2]{TianSunChen18_1sub} are randomly generated
according to a random strategy $\vmat{S}=(S_1,\ldots,S_{\const{M}-1})$ with independent and identically distributed
(i.i.d.) entries according to $\Uniform{[0:n-1]}$.
In this section, we introduce an \rc{$(\const{M},n)$ WPIR scheme, referred to as \emph{Scheme~1}} and denoted by
$\collect{C}_1$, based on the PIR scheme in~\cite{TianSunChen18_1sub}. Scheme~1 can be seen as a generalization of the
PIR scheme in~\cite{TianSunChen18_1sub} where we lift the perfect privacy condition~\eqref{eq:strong-privacy}.


For the proposed scheme, we assume the file size to be $\beta=n-1$, and we represent a query by a length-$\const{M}$
vector $\mat{q}_l=(q_{l,1},\ldots,q_{l,\const{M}})\in\set{Q}_l\subseteq[0:n-1]^{\const{M}}$. Also, the realization of
$\vmat{S}$ is denoted as a length-$(\const{M}-1)$ vector $\vect{s}=(s_1,\ldots,s_{\const{M}-1})$, $s_j\in [0:n-1]$,
$j\in [\const{M}-1]$.

Before describing Scheme~1 in detail for the general case, for simplicity we first present Scheme~1 for the case of
$\const{M}=2$ files and $n=2$ servers (i.e., both servers $1$ and $2$ store $\vmat{X}^{(1)}$, $\vmat{X}^{(2)}$) in the
following example.
\begin{example}
  We illustrate \rc{the} $(2,2)$ Scheme~1 obtained by adopting a nonuniformly-distributed random strategy $\vmat{S}$
  giving a joint PMF $P_{\vmat{Q}_1,\vmat{Q}_2}(\vmat{q}_1,\vmat{q}_2)$ as
    \begin{IEEEeqnarray*}{rCl}
   \Scale[0.9]{ P_{\vmat{Q}_1^{(1)},\vmat{Q}_2^{(1)}}(\mat{q}_1,\mat{q}_2)}& = &\Scale[0.9]{
    \begin{cases}
      1-p & \textnormal{if } \mat{q}_1=(0,0),\mat{q}_2=(1,0),
      \\
      p & \textnormal{if } \mat{q}_1=(1,1),\mat{q}_2=(0,1),
      \\
      0 & \textnormal{otherwise},
    \end{cases}}\\
    \Scale[0.9]{P_{\vmat{Q}_1^{(2)},\vmat{Q}_2^{(2)}}(\mat{q}_1,\mat{q}_2)}& = &\Scale[0.9]{
    \begin{cases}
      1-p & \textnormal{if } \mat{q}_1=(0,0),\mat{q}_2=(0,1),
      \\
      p & \textnormal{if } \mat{q}_1=(1,1),\mat{q}_2=(1,0),
      \\
      0 & \textnormal{otherwise}.
    \end{cases}}
  \end{IEEEeqnarray*}
  Files $\vmat{X}^{(1)}$ and $\vmat{X}^{(2)}$ are composed of one stripe each ($\beta=n-1=1$). 
  The answers $\vmat{A}_1$ and $\vmat{A}_2$ are given
  by
  $\bigl(\vmat{A}_1(\mat{q}_1),
  \vmat{A}_2(\mat{q}_2)\bigr)=\bigl(X^{(1)}_{q_{1,1}}+X^{(2)}_{q_{1,2}},X^{(1)}_{q_{2,1}}+X^{(2)}_{q_{2,2}}\bigr)$, where $X^{(m)}_0 = 0$ for all $m \in [2]$.

  One can easily verify that perfect retrievability is satisfied for the above $(2,2)$ IR scheme. Its IR rate is a
  function of $p$ and is given by $\const{R}(p)=\inv{(p+(1-p)+p)}=\inv{(1+p)}$. Observe that $M\indep \vmat{Q}_1$, which implies that $\eMI{M}{\vmat{Q}_1}=\II^{\textnormal{worst}}(M;\vmat{Q}_1)=0$.

  The information leakage is $\rho^{(\mathsf{MI})}=\frac{1-\Hb(p)}{2}$ and $\rho^{(\textnormal{WIL})}=1-\Hb(p)$ for
  $0\leq p\leq\frac{1}{2}$, where $\Hb(p)\eqdef -p\log_2{p}-(1-p)\log_2{(1-p)}$ is the binary entropy function. From
  this \rc{derivation}, it follows that the  $(2,2)$ Scheme~1 achieves perfect privacy for $p=\frac{1}{2}$. The IR rate 
  of the $(2,2)$ Scheme~1, $\const{R}(\collect{C}_1)$, is depicted in Fig.~\ref{fig:WPIRn2_M2} as a function of the information leakage
  $\rho^{(\cdot)}$. Interestingly, by sacrificing perfect privacy, it is possible to achieve an IR rate larger than the
  $2$-server PIR capacity for $2$ files. As expected, the IR rate increases with increasing information leakage.
  \begin{figure}[t!]
    \centering
    \includegraphics[width=\columnwidth,height=5.5cm,keepaspectratio]{\Figs/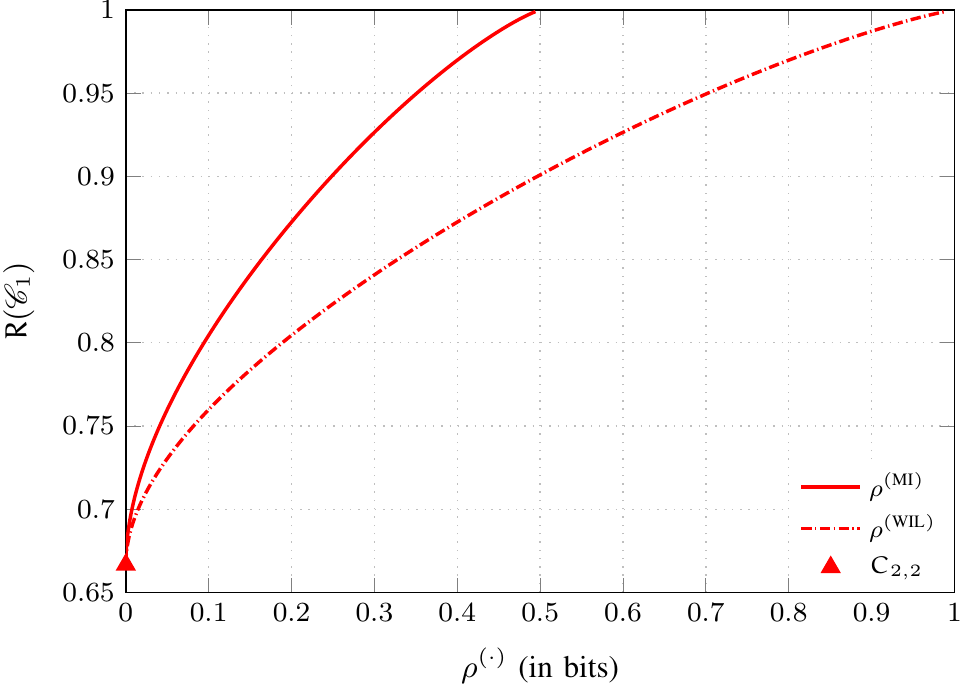}
    \vspace{-1ex}
    \caption{The IR rate $\const{R}(\collect{C}_1)\in\bigl[\frac{2}{3},1\bigr]$ of the proposed $(2,2)$ Scheme~1, as a
      function of $\rho^{(\cdot)}$. The triangle marks the $2$-server PIR capacity for $\const{M}=2$.}
    \label{fig:WPIRn2_M2}
    \vspace{-2ex}
  \end{figure}  
\end{example}



Now, we describe Scheme~1 for the general case of $\cM$ files and $n$ servers. We assume that the user
wants to download file $\vmat{X}^{(m)}$ and \rc{has} a random strategy $\vmat{S}$ that takes on values
$\mat{s}\in [0:n-1]^{\const{M}-1}$ with PMF $P_{\vmat{S}}(\mat{s})$.

\subsubsection{Query Generation}
\label{sec:query-generation_Scheme1}

The query $\mat{q}_l\in\set{Q}_{l}$ sent to the $l\textnormal{-th}$ server, resulting from the query-encoding function
$\phi_l$, is defined as $\mat{q}_l=\bigl(s_1,\ldots,s_{m-1},q_{l,m},s_{m},\ldots,s_{\const{M}-1}\bigr)$, where
$q_{l,m}=\bigl(l-1-\sum_{j\in[\const{M}-1]}s_j\bigr)\bmod n$. It follows \rc{that}
$\bigl(\sum_{m'\in[\const{M}]}q_{l,m'}\bigr)\bmod n=l-1$. Note that the PMF of $\vmat{Q}_l$ conditioned on the file
index $M$ satisfies $P_{\bm Q^{(m)}_l}(\bm q_l)=P_{\vmat{S}}(\mat{s})$.

\subsubsection{Answer Construction}
\label{sec:answer-construction_Scheme1}

The answer function $\varphi_l$ maps  the query $\mat{q}_l$ into
$\vmat{A}_l=\varphi_l(\mat{q}_l,\vmat{X}^{[\const{M}]})=X^{(1)}_{q_{l,1}}+\cdots+X^{(\const{M})}_{q_{l,\const{M}}}$, where $X^{(m')}_0 = 0$ for all $m' \in [\const{M}]$. Further, we see that the
answer-length functions satisfy
\begin{IEEEeqnarray}{c}
  L_l(\mat{A}_l)=
  \begin{cases}
    0 & \textnormal{if } \mat{q}_l=\vect{0},
    \\
    1 & \textnormal{otherwise}.
  \end{cases}
  \label{eq:Ll_Scheme1}
\end{IEEEeqnarray}

This completes the construction of the $(\const{M},n)$ Scheme~1. The perfect retrievability of Scheme~1 can be verified
by following the same argumentation as in \cite[Sec.~3.2]{TianSunChen18_1sub}.
Moreover, using \eqref{eq:Ll_Scheme1}, the IR rate of the $(\cM,n)$ Scheme~1, $\collect{C}_1$, can be shown to be
\begin{IEEEeqnarray*}{c}
  \const{R}(\collect{C}_1)
  =\frac{n-1}{1-P_{\vmat{Q}_1}(\vect{0})+n-1}.
\end{IEEEeqnarray*}
We also remark that if Scheme~1 uses a random strategy $\vmat{S}$ with $\{S_j\}_{j=1}^{\const{M}-1}$ i.i.d.\ according
to $\Uniform{[0:n-1]}$, then it satisfies \eqref{eq:strong-privacy} and is equivalent to the PIR capacity-achieving
scheme proposed in \cite{TianSunChen18_1sub}.

\subsection{$(\const{M},2)$ Scheme~1 With $\{S_j\}_{j=1}^{\const{M}-1}$  I.I.D.\ According to $\Bernoulli{p}$}
\label{sec:achievability_Mn2-Scheme1-IID-Bernoulli}

The following result gives an achievable $4$-tuple for Scheme~1 for the case of $2$ servers and a random strategy
$\vmat{S}=(S_1,\ldots,S_{\const{M}-1})$ with i.i.d.\ entries according to $\Bernoulli{p}$.
\begin{theorem}
  \label{thm:Scheme1_Mn2-IID-Bernoulli} 
  Consider $0\leq p\leq\frac{1}{2}$. Then, the $4$-tuple $(\const{R}_1,\const{U}_1,\Delta_1,\rho^{(\cdot)}_1\bigr)$,
  \begin{IEEEeqnarray*}{rCl}
    \const{R}_1& = &\inv{\bigl(1-(1-p)^{\const{M}-1}+1\bigr)},
    \\
    \const{U}_1& = &\Scale[0.95]{-\sum_{w=0}^\const{M}{\const{M}\choose w} f(w,p)\log_2 f(w,p) },
    \\
    \Delta_1& = &\Scale[0.95]{\sum_{w=0}^\const{M}w\binom{\const{M}}{w} f(w,p)},
    \\
    \rho^{(\mathsf{MI})}_1& = &\const{U}_1/2-(\const{M}-1)\Hb(p), \textnormal{ and}
    \\
    \rho^{(\mathsf{WIL})}_1& = &\log_2{\const{M}}-\min\nolimits_{w\in [0:\const{M}]}\eHP{M_w},
  \end{IEEEeqnarray*}
  where $f(w,p)\eqdef\bigl[(\const{M}-w)(1-p)^{\const{M}-w-1}p^w+ w(1-p)^{\const{M}-w}p^{w-1}\bigr]/\const{M}$ and $M_w$
  has PMF
  \begin{IEEEeqnarray*}{c}
    P_{M_w}(m')=
    \begin{cases}
      \frac{(1-p)^{\const{M}-w-1}p^{w}}{\const{M}f(w,p)} & \textnormal{if }m'\in[\const{M}-w],
      \\[2mm]
      \frac{(1-p)^{\const{M}-w}p^{w-1}}{\const{M}f(w,p)} & \textnormal{if }m'\in[\const{M}-w+1:\const{M}],
    \end{cases}
    \IEEEeqnarraynumspace
  \end{IEEEeqnarray*}
  is achievable by the $(\const{M},2)$ Scheme~1 with $\{S_j\}_{j=1}^{\const{M}-1}$ i.i.d.\ according to $\Bernoulli{p}$.
\end{theorem}

\subsection{Partition Scheme~1: Using Scheme~1 as a Subscheme}
\label{sec:partition-SchemeA}

In Section~\ref{sec:partition-schemes}, the concept of adopting an existing $(\const{M}/\eta,n)$ IR scheme to retrieve a
file from a given partition is introduced. In this section, unlike \eqref{eq:queries_partition-scheme}, where the user
sends different queries for different requested files among all partitions, we use a slightly more sophisticated way to
construct a WPIR scheme by using Scheme~1 as a subscheme for every partition. We refer to this scheme as
\rc{\emph{partition Scheme~1}} and denote it by $\collect{C}_1^\mathsf{part}$. In the following, we present the query
generation and the answer construction.

\subsubsection{Query Generation}
\label{sec:query-generation_SchemeA}

We consider the $j$-th partition, $\set{P}_j$, $j\in[\eta]$, containing \rc{all files of} indices
$(j-1)\const{M}/\eta+1,\ldots,j\const{M}/\eta$. Given a requested file with index
$m=(j-1)\const{M}/\eta+m'\in\set{P}_j$, $m'\in[\const{M}/\eta]$, we consider an $(\const{M}/\eta,n)$ Scheme~1 as a
subscheme for partition $\set{P}_j$. The $l$-th query $\mat{q}_l\in\set{Q}_l$, $l\in[n]$, is defined as
\begin{IEEEeqnarray*}{rCl}
  \mat{q}_l& = &\Scale[0.95]{\bigl(\vect{0}_{1\times
      (j-1)\const{M}/\eta},s_{1},\ldots,s_{m'-1},q_{l,(j-1)\const{M}/\eta+m'},}
  \nonumber\\
  && \quad\>\Scale[0.95]{s_{m'},\ldots,s_{\const{M}/\eta-1},\vect{0}_{1\times(\eta-j)\const{M}/\eta}\bigr)},
\end{IEEEeqnarray*}
where $q_{l,(j-1)\const{M}/\eta+m'}=\bigl(l-1-\sum_{j\in[\const{M}/\eta-1]} s_j \bigr)\bmod n$. We remark that it is
possible that the user sends the all-zero query $\mat{q}_l=\vect{0}$ to request different files among all partitions. In
this way, since the uncertainty on the requested file is increased, it follows that the leakage of $\collect{C}_1^\mathsf{part}$ is slightly smaller than the leakage of the basic
scheme. Moreover, the query alphabet size is not exactly the same for all servers, i.e., we have
$\card{\set{Q}_1}=1+\eta\bigl(n^{\const{M}/\eta-1}-1\bigr)$ and $\card{\set{Q}_l}=\eta\cdot n^{\const{M}/\eta-1}$ for
$l\in [2:n]$.

\subsubsection{Answer Construction}
\label{sec:answer-construction_SchemeA}

Similar to Scheme~1 in Section~\ref{sec:Scheme1}, the answer function $\varphi_l$  maps  query
$\mat{q}_l$ into $\vmat{A}_l=\varphi(\mat{q}_l,\vmat{X}^{[\const{M}]})=X^{(1)}_{q_{l,1}}+\cdots+X^{(\const{M})}_{q_{l,\const{M}}}$, 
where $X^{(m')}_0 = 0$ for all $m' \in [\const{M}]$. 
Further, we see that
$L_l(\vmat{A}_l)$ satisfies~\eqref{eq:Ll_Scheme1}.


\subsection{$(\const{M},2)$ Partition Scheme~1 With $\{S_j\}_{j=1}^{\const{M}/\eta-1}$ I.I.D.\ According to
  $\Bernoulli{1/2}$}
\label{sec:achievability_Mn2-SchemeA-IID-Uniform}

We focus again on the case of $2$ servers. Since the servers can learn some information from which partition the
requested file belongs to, in order to have a relatively small leakage of partition Scheme~1, it is reasonable to use
Scheme~1 with $\{S_j\}_{j=1}^{\const{M}/\eta-1}$ i.i.d.\ according to $\Bernoulli{1/2}$ as a subscheme (i.e., a PIR
subscheme). We have the following result.
\begin{theorem}
  \label{thm:SchemeA_Mn2-IID-Uniform}
  Let $\const{M}/\eta$ be a positive integer with $\eta\in [\const{M}-1]$. Then, the $4$-tuple
  $\bigl(\const{R}_{1,\mathsf{P}},\const{U}_{1,\mathsf{P}},\Delta_{1,\mathsf{P}},\rho^{(\cdot)}_{1,\mathsf{P}}\bigr)$,
  \begin{IEEEeqnarray*}{rCl}
    \const{R}_{1,\mathsf{P}}& = &\inv{\bigl(1+1/2+\cdots+1/2^{\const{M}/\eta-1}\bigr)},
    \\
    \const{U}_{1,\mathsf{P}}& = &\Scale[0.95]{2\bigl[\const{M}/\eta-1+\log_2{\eta}\bigr]
      -\log_2{\eta}/{2^{\const{M}/\eta-1}}},
    \\
    \Delta_{1,\mathsf{P}}& = &\const{M}/\eta,
    \\[1mm]
    \rho^{(\mathsf{MI})}_{1,\mathsf{P}}& = &\log_2{\eta}-\log_2{\eta}/2^{\const{M}/\eta}, \textnormal{ and}
    \\    
    \rho^{(\mathsf{WIL})}_{1,\mathsf{P}}& = &\log_2{\eta}
  \end{IEEEeqnarray*}
  is achievable by the $(\const{M},2)$ partition Scheme~1 with the $(\const{M}/\eta,2)$ Scheme~1 with
  $\{S_j\}_{j=1}^{\const{M}/\eta-1}$ i.i.d.\ according to $\Bernoulli{1/2}$ as a subscheme.
\end{theorem}

Let $\bigl(\tilde{\const{R}},\tilde{\const{U}},\tilde{\Delta},0\bigr)$ be the achievable $4$-tuple of the
$(\const{M}/\eta,2)$ Scheme~1 with $\{S_j\}_{j=1}^{\const{M}/\eta-1}$ i.i.d.\ according to $\Bernoulli{1/2}$. It follows
that
$\const{U}_{1,\mathsf{P}}=\tilde{\const{U}}+2\log_2{\eta}-\log_2{\eta}/2^{\const{M}/\eta-1}
<\const{U}(\collect{C}^{\mathsf{basic}})$ and
$\rho^{(\mathsf{MI})}_{1,\mathsf{P}}=\log_2{\eta}-\log_2{\eta}/2^{\const{M}/\eta}
<\rho^{(\mathsf{MI})}(\collect{C}^{\mathsf{basic}})$, while $\const{R}_{1,\mathsf{P}}$, $\Delta_{1,\mathsf{P}}$, and
$\rho^{(\mathsf{WIL})}_{1,\mathsf{P}}$ are identical to those of the basic scheme $\collect{C}^{\mathsf{basic}}$ in
Section~\ref{sec:partition-schemes} (see the details in Theorem~\ref{thm:partition-schemes}).
Hence, in the numerical results section (see Fig.~\ref{fig:tuple-MI_n2M32}), the results of
$\collect{C}^{\mathsf{basic}}$ are not presented.

\section{Constant-Rate $(\const{M},n)$ Scheme~2}
\label{sec:Scheme2}

We propose an alternative WPIR scheme, referred to as \rc{\emph{Scheme~2} and} denoted by $\collect{C}_2$, based on the
PIR scheme in \cite[Lem.~4]{KumarLinRosnesGraellAmat19_1app}. Scheme~2 is constructed as follows. Assume that
$\beta=n-1$ and that the user requests file $\vmat{X}^{(m)}$. The random strategy $\vmat{S}$ takes the form of a vector
$\vmat{S}=(S_1,\ldots,S_{\beta\const{M}})\in\GF(q)^{\beta\const{M}}$ of length $\beta\const{M}$. The query vector
$\vmat{Q}_l\in\set{Q}_l=\GF(q)^{\beta\const{M}}$, of length $\beta\const{M}$, is obtained as
$\vmat{Q}_l=\phi(m,\bm S)=\bm S+\bm v_l$, where \rc{the} vector $\mat{v}_l$ is deterministic and \rc{is} completely
determined by $m$. We refer the reader to
\cite[Sec.~V]{KumarLinRosnesGraellAmat19_1app} for details on the design of $\mat{v}_l$. The $l$-th server responds to its
corresponding query with the answer $A_l\in\set{A}=\GF(q)$ obtained as
$A_l=\varphi(\vmat{Q}_l,\vmat{X}^{[\const{M}]})=
\langle\vmat{Q}_l,(X^{(1)}_1,\ldots,X^{(1)}_\beta,\ldots,X^{(\const{M})}_\beta)\rangle$.

For the case where $\{S_j\}_{j=1}^{\beta \const{M}}$ are i.i.d. according to $\Uniform{\text{GF}(q)}$, Scheme~2 achieves
perfect privacy, and the scheme boils down to the PIR scheme in
\cite[Lem.~4]{KumarLinRosnesGraellAmat19_1app}. Furthermore, similar to \cite[Thm.~2]{KumarLinRosnesGraellAmat19_1app},
it can be shown that the scheme achieves perfect retrievability (see \eqref{eq:retrievability})\lin{, and since its
  answer-lengths are constant for all possible queries of each server, the IR rate $\const{R}_2$ of $\collect{C}_2$ is
  equal to $1-1/n$, irrespective of the information leakage $\rho^{(\cdot)}$.}

\subsection{$(\const{M},2)$ Scheme~2  With $\{S_j\}_{j=1}^{\const{M}}$ I.I.D.\ According to $\Bernoulli{p}$}
\label{sec:achievability_Mn2-Scheme2-IID-Bernoulli}

\lin{Consider the binary field.} We have the following result.
\begin{theorem}
  \label{thm:Scheme2_n2infnity}
  Consider $0\leq p\leq\frac{1}{2}$. Then, the $4$-tuple $\bigl(1/2,\const{U}_2,\Delta_2,\rho^{(\cdot)}_2\bigr)$,
  \begin{IEEEeqnarray*}{rCl}
    \const{U}_2& = &\Scale[0.95]{-\sum_{w=0}^{\const{M}}\binom{\const{M}}{w}g(w,p)\log_2 g(w,p)+\const{M}\Hb(p)},
    \\
    \Delta_2& = &\Scale[0.95]{\sum_{w=0}^{\const{M}}w\binom{\const{M}}{w}\bigl(g(w,p)+h(w,p)\bigr)},
    \\
    \rho_2^{\mathsf{(MI)}}& = &\Scale[0.95]{\const{U}_2/2-\const{M}\Hb(p)},  \textnormal{ and}
    \\
    \rho_2^{\mathsf{(WIL)}}& = &\Scale[0.95]{\log_2{\const{M}}-\min_{w\in[0:\const{M}]}\eHP{M_w'}},
  \end{IEEEeqnarray*}
  where
  $g(w,p)\eqdef\frac{1}{\const{M}}\bigl[(\const{M}-w)(1-p)^{\const{M}-w-1}p^{w+1}+w(1-p)^{\const{M}-w+1}p^{w-1}\bigr]$,
  $h(w,p)\eqdef(1-p)^{\const{M}-w}p^{w}$, and $M_w'$ has PMF
  \begin{IEEEeqnarray*}{c}
    P_{M_w'}(m')=
    \begin{cases}
      \frac{(1-p)^{\const{M}-w-1}p^{w+1}}{\const{M}g(w,p)} & \textnormal{if } m'\in[\const{M}-w],
      \\
      \frac{(1-p)^{\const{M}-w+1}p^{w-1}}{\const{M}g(w,p)} & \textnormal{if } m'\in[\const{M}-w+1:\const{M}],
    \end{cases}
    \IEEEeqnarraynumspace
  \end{IEEEeqnarray*}
  is achievable by the $(\const{M},2)$ Scheme~2 with $\{S_j\}_{j=1}^{\const{M}}$ i.i.d. according to $\Bernoulli{p}$.
\end{theorem}

In the following subsection, we analyze the $(\const{M},2)$ Scheme~2 with a uniformly-distributed random strategy
$\vmat{S}$. Note that \rc{similarly} to partition Scheme~1 in Section~\ref{sec:achievability_Mn2-SchemeA-IID-Uniform},
we can also construct a partition scheme by using Scheme~2 as a subscheme for every partition. Since the analysis is
almost the same as for partition Scheme~1, and the result for the $(\const{M},2)$ partition Scheme~2 with
$\{S_j\}_{j=1}^{\const{M}/\eta}$ i.i.d.\ according to $\Bernoulli{1/2}$ is very close to the result in
Theorem~\ref{thm:SchemeA_Mn2-IID-Uniform}, we omit it.

\subsection[\texorpdfstring{Math symbols $S$}{Math symbols S}]{$(\const{M},2)$ Scheme~2 With $\vmat{S}$ Uniformly
  Distributed}
\label{sec:achievability_Mn2-Scheme2-UniformSphere}

We consider the $(\const{M},2)$ Scheme~2 with $\vmat{S}$ uniformly distributed over all length-$\const{M}$ binary
vectors of weight $w$. In other words, $\vmat{S}\sim\Uniform{\set{B}_{w,\const{M}}}$, where
$\set{B}_{w,\const{M}}\eqdef\{\mat{s}\in\{0,1\}^\const{M}\colon w_\mathsf{H}(\mat{s})=w\}$.
\begin{theorem}
  \label{thm:Scheme2_Mn2-UniformSphere}
  Consider $0\leq w\leq\const{M}$. Then, the $4$-tuple
  $\bigl(1/2,\const{U}_{2,\mathsf{U}},\Delta_{2,\mathsf{U}},\rho^{(\cdot)}_{2,\mathsf{U}}\bigr)$,
  \begin{align*}
    \const{U}_{2,\mathsf{U}}& =\Scale[0.95]{\log_2\binom{\const{M}}{w}+y(w,\const{M})},
    \\
    \Delta_{2,\mathsf{U}}& =1+2w(1-1/\const{M}),
    \\[1mm]
    \rho^{(\textnormal{MI})}_{2,\mathsf{U}}& =\Scale[0.95]{\bigl(y(w,\const{M})-\log_2{\binom{\const{M}}{w}}\bigr)/2},
                                             \textnormal{ and}
    \\
    \rho^{(\textnormal{WIL})}_{2,\mathsf{U}}& =\Scale[0.95]{\log_2{\const{M}}-\min\{\log_2{(w+1)},\log_2{(\const{M}-w+1)}\}}, 
  \end{align*}
  where
  $y(w,\const{M})\eqdef\log_2{\binom{\const{M}}{w}}+\log_2{\const{M}}-\frac{\const{M}-w}{\const{M}}\log_2{(w+1)}
  -\frac{w}{\const{M}}\log_2{(\const{M}-w+1)}$, is achievable by the $(\const{M},2)$ Scheme~2 with
  $\vmat{S}\sim\Uniform{\set{B}_{w,\const{M}}}$.
\end{theorem}

We remark that the analysis of the $(\const{M},2)$ Scheme~1 with $\vmat{S}\sim\Uniform{\set{B}_{w,\const{M}-1}}$
    can also be
done by following the same approach as for Theorem~\ref{thm:Scheme2_Mn2-UniformSphere}.\footnote{\lin{The scheme is not equal to that of \cite{TianSunChen18_1sub} because of the difference in the vector space of the random strategy. The former involves all length-($\const{M}-1$) vectors of weight $w$, while the latter consists of all vectors of length $\const{M}-1$.}} However, since the resulting
performance is much worse than those of the aforementioned WPIR schemes for the case of $n=2$ servers, we omit the
detailed analysis
in this paper.

\section{Numerical Results}
\label{sec:numerical-results}

We consider the case of $2$ servers and compare the achievable $4$-tuples
$\bigl(\const{R},\const{U},\Delta,\rho^{(\M)}\bigr)$ for the $(\const{M},2)$ IR schemes proposed in
\cref{sec:achievability_Mn2-Scheme1-IID-Bernoulli,sec:achievability_Mn2-SchemeA-IID-Uniform,sec:achievability_Mn2-Scheme2-IID-Bernoulli,sec:achievability_Mn2-Scheme2-UniformSphere}. For
the sake of illustration, the information leakage $\rho^{(\M)}$ is normalized by $\log_2{\const{M}}$ bits, while the
upload cost and access complexity are normalized by $2(\const{M}-1)$ and $\const{M}$, respectively. $2(\const{M}-1)$ and
$\const{M}$ are the upload cost and access complexity of the PIR capacity-achieving scheme presented in
\cite{TianSunChen18_1sub} for the case of $2$ servers. The upload cost $2(\const{M}-1)$ is optimal among all so-called
\emph{decomposable} PIR capacity-achieving schemes \cite{TianSunChen18_1sub}.\footnote{Based on
  \cite[Def.~2]{TianSunChen18_1sub}, all existing PIR schemes in the literature are decomposable.}

\begin{figure}[t]
  \centering
  \includegraphics[width=\columnwidth,,height=11cm,keepaspectratio]{\Figs/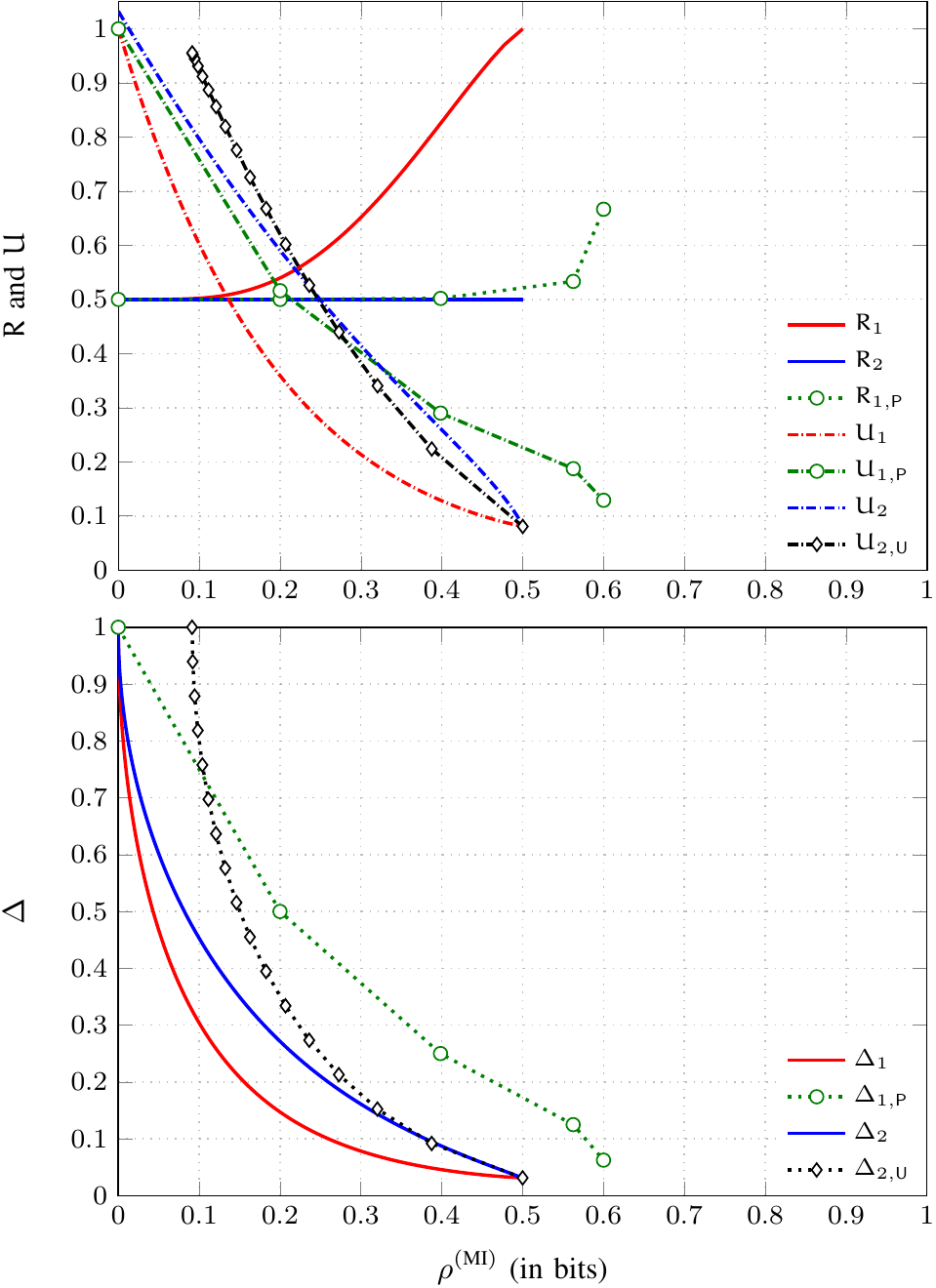}
  \vspace{-1ex}
  \caption{$\const{R}$, $\const{U}$, and $\Delta$ of different WPIR schemes for $\const{M}=32$, as a function of
    $\rho^{(\M)}$. For $\const{M}=32$, $\const{C}_{\const{M},2}$ is almost equal to $1/2$.}
  \label{fig:tuple-MI_n2M32}
  \vspace{-4ex}
\end{figure}

Fig.~\ref{fig:tuple-MI_n2M32} illustrates the results of different WPIR schemes for the case of $\const{M}=32$ files and
leakage metric $\rho^{(\mathsf{MI})}$. We can see that Scheme~1 yields the best performance. The IR rate of Scheme~2
with different $\vmat{S}$ is always equal to $1/2$. The results of different WPIR schemes in terms of leakage metric
\rc{$\rho^{(\mathsf{WIL})}$} will be provided in the extended version of the paper.

\section{Conclusion}
\label{sec:conclusion}

We presented the first study of IR schemes with information leakage, which we refer to as WPIR schemes. We proposed two
WPIR schemes based on two different PIR protocols and a family of schemes based on partitioning for the case of
replication. By relaxing the perfect privacy requirement, we showed that the download rate, the upload cost, and the
access complexity can be improved.

\bibliographystyle{IEEEtran}
\bibliography{./defshort1,./biblioHY}

\end{document}